\documentclass{article}
\usepackage{amsmath,amssymb}
\usepackage{graphicx}
\numberwithin{equation}{section}

\begin{document}

\title{Slowly rotating charged fluid balls in the presence of a cosmological constant}

\author{
Matthew Wright\footnote{matthew.wright.13@ucl.ac.uk}\\
Department of Mathematics, University College London\\
Gower Street, London, WC1E 6BT, UK
}

\date{\today}

\maketitle

\begin{abstract}
We examine charged slowly rotating perfect fluids in the presence of a cosmological constant. The asymptotic form of the vacuum solutions to the linearised Einstein-Maxwell field equations is found and the possibility of matching this vacuum to the slow rotating Garc\'{\i}a metric is considered. We show that, contrary to the case of zero cosmological constant, this Garc\'{\i}a metric can  be matched to an asymptotically de Sitter vacuum in the slow rotation limit. We conclude the Garc\'{\i}a metric may potentially be suitable for describing a charged isolated rotating body in a cosmological background.
\end{abstract}

\section{Introduction}

It is remarkable that one hundred years after the general theory of relativity was first formulated that there are still no known exact analytical solutions to Einstein's equations describing an isolated rotating body.  There are of course highly accurate numerical solutions of rotating stars~\cite{Stergioulas:2003yp,Stergioulas:2013}, but mathematically it is of interest to obtain analytic solutions. In particular, and of main interest in this paper, there are no exact solutions describing an interior rotating charged perfect fluid which can be matched to an asymptotically flat vacuum exterior. One reason for this is the sparsity of known analytic charged rotating fluid interiors. One such interior is the Garc\'{\i}a solution~\cite{Garcia:1991}, also known as the Wahlquist-Newman metric~\cite{mars}, which is the charged generalisation of the rotating Wahlquist solution~\cite{Wahlquist:1968zz}. The Wahlquist solution is in turn the rotating generalisation of the static Whittaker metric. These metrics all have unphysical equations of state, in the Wahlquist and Whittaker case the simple relation $\mu+3p=\text{const}$ holds, where $\mu$ is the energy density and $p$ is the pressure. Nonetheless due to the lack of explicit solutions the possibility of matching these solutions to external vacuum domains should be investigated.

The slow rotation formalism developed by Hartle in 1967~\cite{Hartle:1967he} has been important in deriving results about the possibility of matching these interior solutions to exterior vacuum regions. If the matching is not possible to a particular exterior in the slow rotation limit, then it will not be possible for a more rapid rotation. This argument has been used to show that the Wahlquist metric cannot be matched to an asymptotically flat vacuum~\cite{Bradley:1999gs}, and one proves this by expanding the Wahlquist metric to second order in the fluid's angular velocity. If one drops the requirement that the vacuum is asymptotically flat, the matching becomes possible~\cite{Bradley:2000sj}, however this means the Wahlquist metric cannot serve as a model for an isolated body, as a far away quadrupole mass distribution is required to keep the body in equilibrium.

The matching conditions become more restrictive once charge is included. In the absence of charge, if one linearises the Einstein equations in the angular velocity of the metric, one can show that the general slowly rotating exterior vacuum solution of a perfect fluid sphere coincides with the Kerr solution, and one can match any slow rotating perfect fluid interior to this. However in the presence of charge, the general exterior vacuum solution to the linearised Einstein Maxwell equations does not in general coincide with the Kerr-Newman solution, and in fact is in general not asymptotically flat~\cite{Fodor:2002nt}. Thus if one requires asymptotic flatness, as necessary for describing an isolated rotating body in a flat background, then the matching conditions become overdetermined in general.  In the particular case of the Garc\'{\i}a metric, it was shown in~\cite{Fodor:2002nt} that to first order in angular velocity this metric cannot be matched to an asymptotically flat exterior metric. It is found the fluid needs to be embedded in an external magnetic field, parallel to the axes of rotation, and thus the authors conclude that the Garc\'{\i}a metric cannot be suitable for describing an isolated charged body. 

It is now consensus that the universe is accelerating in expansion, which can be well explained by the presence of a small non-zero positive cosmological constant~\cite{Riess:1998cb,Perlmutter:1998np}. Thus the universe is not asymptotically flat, rather asymptotically it behaves like de Sitter space. This observation has resurrected interest in studying such spacetimes, which are either asymptotically de Sitter or anti-de Sitter. Of course, the small observed value of the cosmological constant means that its effects on astrophysical objects are negligible. However the presence of even a tiny cosmological constant completely changes the asymptotic structure of spacetime~\cite{Ashtekar:1984zz,Ashtekar:2014zfa}. Moreover  a given perfect fluid solution to Einstein's equation in the absence of a cosmological constant is also a solution to Einstein's equation with cosmological constant, achieved by making the substitution $p \rightarrow p - \Lambda/(8\pi G)$ for the pressure $p$, and $\mu \rightarrow \mu + \Lambda/(8\pi G)$ for the energy density. This of course changes the equation of state of the fluid, but for the Wahlquist metric, the equation of state retains the same form, $\mu+3p=\text{const}$.  Therefore the possibility of matching rotating fluids to asymptotically (anti)-de Sitter spaces should be investigated. Recently this approach was considered in~\cite{Boehmer:2014kxa}, where it was shown that to second order in the angular velocity the Wahlquist metric can be matched to an asymptotically (anti)-de Sitter external vacuum.  The purpose of this paper is to investigate this possibility for the Garc\'{\i}a metric in the slow rotation limit.

This paper is organised as follows. We begin in Section 2 by reviewing the Garc\'{\i}a solution and its form in the slow rotation limit. In Section 3 we examine the general slow rotating electro-lambda-vacuum exterior metric, which we show is always asymptotically (anti)-de Sitter, with the electromagnetic field tensor likewise asymptotically decaying. Finally in Section 4 we describe the matching procedure and we show that up to first order in the angular velocity one can in fact match the Garc\'{\i}a metric to an asymptotically (anti)-de Sitter vacuum. Thus we conclude by noting that the Garc\'{\i}a metric may be suitable for describing an isolated rotating charged body in (anti)-de Sitter spaces.

\section{The Garc\'{\i}a solution}
In this section we briefly review the Garc\'{\i}a solution and dicuss its form in the slow rotation limit. The Garc\'{\i}a solution, first discovered in 1991~\cite{Garcia:1991}, is the charged generalisation of the Wahlquist metric~\cite{Wahlquist:1968zz} and is given by
\begin{align}
ds^2&=\frac{P}{\Delta}(d\tau+\delta N d\sigma)^2-\frac{Q}{\Delta}(d\tau+\delta M d\sigma)^2\nonumber+\Delta(\frac{dx^2}{P}+\frac{dy^2}{Q}),
\end{align}
where
\begin{align}
\Delta&=M-N ,
\nonumber\\
M&= \frac{1}{k^2}\sinh^2(kx)-\xi_0^2,
\nonumber\\
N&= -\frac{1}{k^2}\sin^2(ky)-\xi_0^2,
\nonumber\\
P&= a+\frac{1}{2k}[2n+x(\alpha+\beta^2)]\sinh(2kx)\nonumber+[b-(g+\beta x)^2]\cosh(2kx),
\nonumber\\
Q&= a-\frac{1}{2k}[2m+y(\alpha+\beta^2)]\sin(2ky)\nonumber+[b+(e+\beta y)^2]\cos(2ky).
\end{align}
In the above $a, b, e, g, k, m, n, \alpha, \beta, \delta$ and $\xi_0$ are eleven constants. Up to diffeomorphism invariance only eight of these are independent parameters~\cite{Fodor:2002nt}.

This metric is a solution to the Einstein-Maxwell equations with a cosmological constant
\begin{align}
 G_{\alpha\beta}+\Lambda g_{\alpha\beta}=8\pi(T_{\alpha\beta}^{(f)}+T_{\alpha\beta}^{(e)}),
 \\
 \nabla_{\beta}F^{\beta\alpha}=J^\alpha, \quad \nabla_{[\alpha}F_{\beta\gamma]}=0,
 \end{align}
where $F^{\mu\nu}$ is the electromagnetic field tensor which can be written in terms of the four potential $A_\mu$ as
\begin{align}
F_{\mu\nu}=\nabla_{\mu}A_{\nu}-\nabla_{\nu}A_{\mu}.
\end{align}
$T_{\alpha\beta}^{(e)}$ is the stress energy tensor of the electromagnetic field
\begin{align}
T_{\alpha\beta}^{(e)}=\frac{1}{4\pi}(F_{\alpha\mu}F_{\beta}{}^{\mu}-\frac{1}{4}g_{\alpha\beta}F^{\mu\nu}F_{\mu\nu}),
\end{align}
and $T_{\alpha\beta}^{(f)}$ is the stress energy tensor of a standard perfect fluid with
\begin{align}
T_{\alpha\beta}^{(f)}=(\mu+p)u_\alpha u_{\beta}+p g_{\alpha\beta},
\end{align}
where $u^{\alpha}$ is the four velocity. 

The pressure $p$ and density $\mu$ of this particular solution are given in terms of the metric components by
\begin{align}
8\pi p&=-\frac{k}{\Delta}(Q-P)+\alpha k^2 +\Sigma +\Lambda,
\\
8\pi \mu &= 3\frac{k}{\Delta}(Q-P)-\alpha k^2 -\Sigma -\Lambda,
\end{align}
where the quantity $\Sigma$ has been defined as
\begin{align}
\Sigma&= -\frac{2\beta k}{\Delta}[(e+\beta y)\sin(2ky)+(g+\beta x)\sinh(2kx)].
\end{align}
The four potential $A$ is given by
\begin{align}
A_{\tau}  &=\frac{\Sigma}
{4k^{2}\beta}, \quad A_{x}   =A_{y}=0, \quad  \ \\
A_{\sigma}  & =-\frac{\delta}{2\Delta k}\left[  \left(  e+\beta y\right)
M\sin(2ky)+\left(  g+\beta x\right)  N\sinh(2kx)\right].
\end{align}

Now we examine its form in the slow rotation limit. One can write the general metric of a slowly rotating perfect fluid sphere in the following form~\cite{Hartle:1967he}
\begin{align}
ds^2=-\mathcal{A}^2 dt^2+\mathcal{B}^2 dr^2 +\mathcal{C}^2[d\theta^2+\sin^2\theta (d\phi-\omega dt)^2] \label{slow}
\end{align}
where an expansion in the angular velocity $\Omega$ of the fluid is made. In this paper we consider only the first order perturbations in $\Omega$. This means that the metric functions $\mathcal{A}$, $\mathcal{B}$ and $\mathcal{C}$ are simply the static non-rotating solutions, to first order in $\Omega$ they remain unmodified.  Rotation is added through the function $\omega$ which is assumed to be of $\mathcal{O}(\Omega)$ (and thus we follow others and slightly abuse notation by neglecting the $\omega^2$ component of the above metric). In~\cite{Fodor:2002nt} coordinate transformations of the Garc\'{\i}a metric were derived which allowed it to be put into the above slowly rotating form. It was found that
\begin{align}
ds^2= -h_1 dt^2+\frac{dz^2}{\gamma^2 h_1}+\frac{\sin^2z}{\gamma^2}[d\theta^2+\sin^2\theta(d\varphi-\omega dt^2)],
\end{align}
where the functions $h_1$ and $\omega$ are given by
\begin{align}
h_1&=1-\beta^2+\frac{1}{\kappa^2}(1-z\cot z)+\beta^2z^2(\cot^2z-1),
\\
\omega&=r_0\frac{\gamma^2}{\sin^2z}(h_1-1)
\end{align}
and $r_0$ is of order $\Omega$ and is taken as the slow rotation parameter. In the static limit $r_0\rightarrow0$ this metric reduces to the charged Whittaker metric. The density and pressure are now given by
\begin{align}
8\pi p&= \gamma^2(-h_1+\frac{1}{\kappa^2}-\beta^2)+4\beta\gamma^2 A_t+\Lambda \nonumber \\
8\pi \mu&=\gamma^2(3h_1-\frac{1}{\kappa^2}+\beta^2)-4\beta\gamma^2 A_t-\Lambda.
\end{align}
We must also expand the four potential $A$ up to first order in $r_0$, and it is found that
\begin{align}
A&=-\beta z \cot z dt\nonumber+r_0[\beta \sin^2\theta(1-z\cot z)-\beta-\bar{g}\cos\theta]d\varphi.
\end{align}
In this expression $\bar{g}$ is the magnetic monopole charge parameter. The metric is completely independent of this parameter, and it only contributes to the Maxwell equations.

The set of constants in the original metric have now been transformed to a new set of five constants $\beta, \kappa, \gamma, \bar{g}$ and $r_0 $. The number of constants has been reduced from the original eight independent parameters (modulo diffeomorphisms) by fixing a coordinate system and ensuring that the metric is completely regular at the center. 

\section{Electrovacuum exterior}

Now in this section we will find the corresponding slow rotating form of the vacuum exterior. We make a slow rotation approximation around Reissner-Nordstr\"{o}m-de Sitter space by expanding the metric up to first order in the angular velocity $\Omega$. We write the metric in the same form as the slow rotating Garc\'{\i}a metric~(\ref{slow}). Now from the Einstein equations, to first order in the angular velocity, the diagonal components of the metric are unmodified and the only modification is to the $g_{t\varphi}$ component, see~\cite{Hartle:1967he}. Hence we may write the electrovacuum exterior as
\begin{align}
ds^2&=-(1-\frac{2M}{r}+\frac{e^2}{r^2}-\frac{\Lambda r^2}{3})dt^2 \nonumber+(1-\frac{2M}{r}+\frac{e^2}{r^2}-\frac{\Lambda r^2}{3})^{-1}dr^2
\nonumber\\&+ r^2[d\theta^2+\sin^2 \theta(d\varphi-\omega dt)^2],
\end{align}
where $M$ is the mass, $e$ is the charge and $\Lambda$ is the cosmological constant, which we consider either to be positive or negative. The function $\omega$ is first order in the angular velocity, and again we ignore the $\omega^2$ component of the above metric. In the Garc\'{\i}a interior solution, the function $\omega$ is dependent on $r$ only, and we will make the additional assumption that this to be true in the exterior metric also.

We can choose a gauge such that the four potential of the electromagnetic field $A_{\alpha}=A_{\alpha}(r,\theta)$, to zero-th order in angular velocity, is
\begin{align}
A_t=-\frac{e}{r}, \quad A_r=0, \quad A_\theta=0, \quad A_\varphi=0.
\end{align}
From the Einstein equations we will see that the only component of the four potential that is perturbed to first order in $\Omega$ is $A_\varphi$. 

\subsection{Slow rotating Kerr-Newman-de Sitter metric}

Let us first examine the slow rotating form of the Kerr-Newman-de Sitter metric. The full metric in Boyer-Lindquist coordinates is given by
\begin{multline}
  ds^2=-\frac{\Delta_l}{\Xi_l^2 \rho_l^2}(dt-a \sin^2\theta d\phi)^2
  +\frac{\Theta_l \sin^2\theta}{\Xi_l^2 \rho_l^2}(adt-(r^2+a^2)d\phi)^2\\
  +\frac{\rho_l^2}{\Delta_l}dr^2+\frac{\rho_l^2}{\Theta_l}d\theta^2,
  \label{kerrbl}
\end{multline}
where
\begin{alignat}{2}
  \rho_l &= \sqrt{r^2+a^2 \cos^2\theta}, &\qquad l&=\sqrt{\frac{3}{\Lambda}}, \\
  \Theta_l&=1+\frac{a^2}{l^2}\cos^2\theta, &\qquad \Xi_l &= 1+\frac{a^2}{l^2},
\end{alignat}
\begin{align}
  \Delta_l=-\frac{r^2}{l^2}(r^2+a^2)+r^2-2Mr+a^2+e^2.
\end{align}
And the electromagnetic potential of this metric is given by
\begin{align}
A=-\frac{er}{\rho_L^2}\left( dt-a \frac{\sin^2\theta}{\Xi_l}d\varphi\right)^2.
\end{align}

To write this in the slow rotating form~(\ref{slow}), we simply expand this metric up to first order in the parameter $a$, which we take to be the rotation parameter and is of order $\Omega$. It is readily seen that the functions $A_{\varphi}$ and $\omega$ are given by
\begin{align}
A_{\varphi}^{\rm KN}= \frac{ae \sin^2\theta}{r}
\end{align} 
\begin{align}
\omega^{\rm KN}=\frac{2aM}{r^3}- \frac{a e^2}{r^4}+\frac{ a\Lambda}{3}.
\end{align}
We are free to perform a rigid rotation of our coordinate system, $\varphi \rightarrow \varphi + c_1 t$, where $c_1$ is a suitably chosen constant to remove the contribution $ a\Lambda/3$ from this expression. This is done so that we are working in an asymptotically non-rotating frame.

\subsection{General vacuum}
Let us now return to working with a general exterior. For a general $\Lambda$-electrovacuum, the $(\varphi,t)$ components of Einstein's equation gives us the following differential equation
\begin{align}
r^4 \frac{d^2 \omega}{d r^2}+4 r^3 \frac{d \omega}{dr}- \frac{4 e}{\sin^2\theta}\frac{\partial A_\varphi}{\partial r}=0.
\end{align}
We can solve this equation to find the general solution of $A_{\varphi}$ for a non-zero charge
\begin{align}
A_\varphi=\frac{1}{4e} r^4 \sin^2\theta \frac{d\omega}{dr}+f(\theta),
\end{align}
where $f$ is an arbitrary function of $\theta$.

We now insert this solution back into the Maxwell equation $\nabla^{\mu}F_{\mu\nu}=0$ and we find the following differential equation relating $\omega$ to $f$
\begin{align}
&\frac{1}{4}(2Mr-e ^2-r^2 +\frac{\Lambda}{3}r^4)r^4\frac{d^3\omega}{dr^3}\nonumber+(\frac{7}{2} M r-2r^2-\frac{3}{2}e^2+\frac{5}{2}\frac{\Lambda}{3}r^4)r^3 \frac{d^2\omega}{dr^2}\nonumber\\&+(4M-\frac{5}{2} r +5 \frac{\Lambda}{3}r^3)r^3 \frac{d\omega}{dr}
 =\frac{e}{\sin^2\theta}(\frac{d^2f(\theta)}{d\theta^2}-\cot\theta\frac{df(\theta)}{d\theta}). \label{ome1eqn}
\end{align}
This is a separable equation, with the left-hand side independent of $\theta$, and hence we may write
\begin{align}
\frac{1}{\sin^2\theta}(\frac{d^2f(\theta)}{d\theta^2}-\cot\theta\frac{df(\theta)}{d\theta})=\frac{K}{e}, \label{feqn}
\end{align}
where $K$ is a constant. We can write the left-hand side of this equation simply as
\begin{align}
\frac{1}{\sin^2\theta}(\frac{d^2f(\theta)}{d\theta^2}-\cot\theta\frac{df(\theta)}{d\theta})=\frac{d^2 f(\cos\theta)}{d (\cos\theta)^2},
\end{align}
so that this equation~(\ref{feqn}) has the simple general solution
\begin{align}
f(\theta)=\frac{1}{2} \frac{K \cos^2 \theta}{e}+\frac{C_4}{e}+\frac{C_5}{e}\cos\theta.
\end{align}

Now examining the radial part of the equation~(\ref{ome1eqn}) gives us a third order inhomogeneous differential equation for $\omega$
\begin{align}
&\frac{1}{4}(2Mr-e ^2-r^2 +\frac{\Lambda}{3}r^4)r^4\frac{d^3\omega}{dr^3}+(\frac{7}{2} M r-2r^2-\frac{3}{2}e^2+\frac{5}{2}\frac{\Lambda}{3}r^4)r^3 \frac{d^2\omega}{dr^2}\nonumber\\&+(4M-\frac{5}{2} r +5 \frac{\Lambda}{3}r^3)r^3 \frac{d\omega}{dr}=K. \label{omegaequation}
\end{align}
The presence of $\Lambda$ in this equation makes finding an analytic solution difficult. However we can find a particular solution, and noting that a constant is also clearly a solution allows us to write the following ansatz for $\omega$
\begin{align}
\omega= C_0 +K\left(\frac{e^2}{3Mr^4}-\frac{2}{3r^3}\right)+C_1\omega_1(r)+C_2\omega_2(r),
\end{align}
where $C_0$, $C_1$ and $C_2$ are all first order constants in $\Omega$ and $\omega_1$ and $\omega_2$ are two linearly independent solutions of~(\ref{omegaequation}). We will analyse the form of these solutions shortly. Now since we know that the slow rotating Kerr-Newman-de Sitter is a particular solution to the above equation, we see that we can set $K=-3aM$, so that taking $C_1=0$, $C_2=0$ we have simply the slow rotating Kerr-Newmann-de Sitter metric. And again, we are always free to set $C_0$ to zero by adjusting to an asymptotically non rotating frame via a rigid rotation of our coordinate system. 

Substituting this solution back into the potential we obtain
\begin{align}
A_{\varphi}&=\frac{ae}{r}\sin^2\theta-\frac{3aM}{2e}+\frac{1}{4e}r^4\sin^2\theta(C_1\omega_1'+C_2\omega_2')+\frac{C_4}{e}+\frac{C_5}{e}\cos\theta.
\end{align}
The term $C_5$ simply gives rise to a magnetic monopole. Now the form of the solutions $\omega_1$ and~$\omega_2$ of~(\ref{omegaequation}) are completely different depending on whether $\Lambda=0$. We investigate the two distinct cases below.

\subsubsection*{Zero cosmological constant $\Lambda=0$}

The solution for $\Lambda=0$ was derived in~\cite{Fodor:2002nt}, and we review its properties here. Assuming $e^2\neq M^2$, it was found 
\begin{align}
\omega_1&=\frac{e^2(r^2-2Mr+e^2)[r+M+(e^2-r^2)L(r)]}{M^2(e^2-M^2)^2r^4}+\frac{2e^2(e^2-2Mr)}{3M^3 r^4(e^2-M^2)} \\
\omega_2&=\frac{2}{r}+\frac{e^2(e^2-2M r)}{Mr^4}
\end{align}
where $L(r)$ is implicitly defined through
\begin{align}
\frac{dL(r)}{dr}=\frac{1}{2Mr-r^2-e^2},
\end{align}
and can be written in terms of $\log$ or $\arctan$ depending on whether the discriminant of the metric component $g_{tt}$, which is given by $4(M^2-e^2)$, is positive or negative respectively. Asymptotically expanding these functions in powers of $1/r$ we find the following far field behaviour
\begin{align}
\omega_1&=\frac{2 e^2}{3 M^3 r^4}+\frac{4 e^2}{5 M^2 r^5}+\ldots  \\
\omega_2&= \frac{2}{r}-\frac{2 e^2}{r^3}+\frac{e^4}{M r^4}+\ldots
\end{align}

Now we can find the asymptotic behaviour of the metric and electromagnetic field tensor by using the following tetrad
\begin{align}
e_0&=(1-\frac{2M}{r}+\frac{e^2}{r^2})^{-1/2}\frac{\partial}{\partial t}\nonumber\\
e_1&=(1-\frac{2M}{r}+\frac{e^2}{r^2})^{1/2}\frac{\partial}{\partial r}\nonumber\\
e_2&=\frac{1}{r}\frac{\partial}{\partial \theta}\nonumber\\
e_3&=\frac{1}{r\sin\theta}\frac{\partial}{\partial \varphi}-\omega r \sin\theta(1-\frac{2M}{r}+\frac{e^2}{r^2})^{-1}\frac{\partial}{\partial t}.
\end{align}
Expressed in this tetrad, it can be shown the electromagnetic field does not vanish asymptotically unless the constant $C_2$ vanishes, since
\begin{align}
\lim\limits_{r\rightarrow\infty}F_{13}&=-\frac{C_2}{e}\sin \theta, \\
\lim\limits_{r\rightarrow\infty}F_{23}&=-\frac{C_2}{e}\cos \theta ,
\end{align}
and similarly by expanding the Weyl curvature tensor in this tetrad, it is also found that the metric is not asymptotically flat unless $C_2=0$. Thus the general vacuum exterior rotating charged metric cannot describe an isolated object. It was shown in~\cite{Fodor:2002nt} that if one attempts to match the Garc\'{\i}a metric to this exterior vacuum, the matching conditions dictate that this constant $C_2\neq0$, and so the exterior vacuum will not be asymptotically flat. Therefore the Garc\'{\i}a metric cannot describe an isolated rotating body in an asymptotically flat spacetime. 

\subsubsection*{Non-zero comological constant $\Lambda\neq 0$}

When $\Lambda\neq 0$ we are unable to find the exact analytic solutions for $\omega_1$ and $\omega_2$ from equation~(\ref{omegaequation}). However we are able to find the form of the solutions asymptotically by making an expansion in powers of $1/r$
\begin{align}
\omega_1&=\frac{1}{r^4}-\frac{6 M}{7 \Lambda 
   r^7}+...\\
 \omega_2&=\frac{1}{r^3}-\frac{9}{5 \Lambda  r^5}-\frac{3 e^2}{7 \Lambda 
       r^7}-\frac{9}{7 \Lambda ^2 r^7}+...
 \end{align}
and we note the function $\omega_1$ now falls off like $1/r^4$ and $\omega_2$ now falls of like $1/r^3$.

To confirm these results, we can also integrate~(\ref{omegaequation}) numerically. We set $K=0$ in the equation so we ignore the particular solution corresponding to the Kerr-Newman-de Sitter contribution and find simply the form of the $\omega_1$ and $\omega_2$ contribution. We also numerically integrate the solution for $\omega'(r)$, since we are only interested in how the solution decays for large $r$, and this will allow us to ignore the constant contribution which arises when numerically integrating $\omega$. 

\begin{figure}
\centering
  \includegraphics[width=0.5\textwidth]{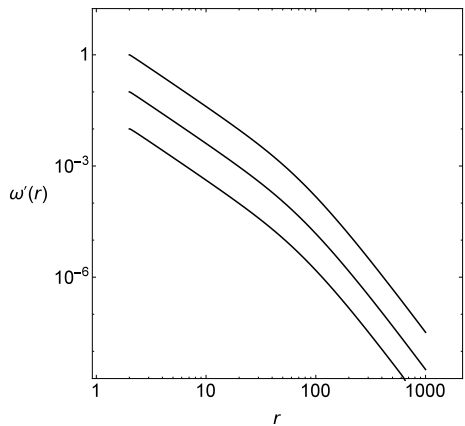}
  \caption{Plot of $\omega'(r)$ against $r$ on a logarithmic scale on both axes. The parameter values are $\lambda = -0.001$ and $e=0.1$ and $M=1$. This means the black hole horizon is located at  $r_{\rm bh } \approx 1.99$. The initial conditions are  $\omega(2)=\{1,0.1,0.01\}$, $\omega'(2)=\{-0.01,0.02,0.01\}$. }
  \label{fig:num1}
\end{figure}

A plot of $\omega'(r)$ against $r$ is plotted on a logarithmic scale in Fig.~(\ref{fig:num1}) for a negative cosmological constant. A positive cosmological constant leads to numerical difficulties at the cosmological horizon, meaning different coordinates would need to be used, however numerically integrating the equation above the horizon gives very much the same result. On logarithmic graphs, the functions $r^n$ are straight lines with gradient $n$, so analysing the gradient of the line on the logarithmic scale allows us to measure the power at which $r$ decays. We find for all three graphs $n \approxeq -3.9$, which supports our asymptotic result that $\omega'$ decays like $1/r^{-4}$ for large $r$.

Now we must investigate whether this observed asymptotic behaviour means that the metric is asymptotically (anti)-de Sitter, and we must also investigate the asymptotic behaviour of the electromagnetic field tensor. This time we use the tetrad
\begin{align}
e_0&=\left(|1-\frac{2M}{r}+\frac{e^2}{r^2}-\frac{\Lambda r^2}{3}|\right)^{-1/2}\frac{\partial}{\partial t}\nonumber\\
e_1&=\left(|1-\frac{2M}{r}+\frac{e^2}{r^2}-\frac{\Lambda r^2}{3}|\right)^{1/2}\frac{\partial}{\partial r}\nonumber\\
e_2&=\frac{1}{r}\frac{\partial}{\partial \theta}\nonumber\\
e_3&=\frac{1}{r\sin\theta}\frac{\partial}{\partial \varphi}-\omega r \sin\theta(1-\frac{2M}{r}+\frac{e^2}{r^2}-\frac{\Lambda r^2}{3})^{-1}\frac{\partial}{\partial t}.
\end{align}
In this tetrad, we obtain the following behaviour of the electromagnetic field asymptotically 
\begin{align}
\lim\limits_{r\rightarrow\infty}F_{13}&=0,\\
\lim\limits_{r\rightarrow\infty}F_{23}&=0,
\end{align}
and in fact these components decay like $1/r^2$, exactly the same as Kerr-Newman-de Sitter space. This is in stark contrast to the $\Lambda=0$ case. 

It is also readily seen that this exterior metric is always asymptotically (anti)-de Sitter. We can see this by examining the asymptotic behaviour of the Weyl tensor. We use the standard definition for the gravitoelectric fields $E_i$ and the gravitomagnetic fields $H_i$ 
\begin{align}
  E_1 & = R_{0101}-\frac{\Lambda}{3}, & \qquad 
  E_2 & = R_{0202}-\frac{\Lambda}{3}, & \qquad 
  E_3 & =R_{0102},\\
  H_1 & = R_{0123}, & \qquad 
  H_2 & = -R_{0213}, & 
  \qquad H_3 &= R_{0223},
\end{align}
with the remaining four components of the Weyl tensor vanishing identically. Using the asymptotic expansion for $\omega$, we find the following expansions for large~$r$
\begin{align}
  E_1 = -\frac{M}{r^3}+\frac{e^2}{r^4}+\mathcal{O}(r^{-5},\Omega^2),
\nonumber\\
  E_2 =    \mathcal{O}(r^{-5},\Omega^2), 
 \quad  E_3 = 
  \mathcal{O}(\Omega^2).
\end{align}
For the gravitomagnetic fields $H_i$ it is found
\begin{align}
  H_1 & = -\frac{3C_2 \cos\theta}{2r^4}+\mathcal{O}(r^{-5},\Omega^2), \qquad 
  H_2  =\frac{3C_2 \cos\theta}{2r^4}+ \mathcal{O}(r^{-5},\Omega^2),
  \nonumber\\
  H_3 & = -\frac{\sqrt{3|\Lambda|}C_2 \sin\theta }{2r^3} -\frac{2\sqrt{3|\Lambda|}C_1 \sin\theta }{3r^3} + 
  \mathcal{O}(r^{-4},\Omega^2).
\end{align}
which decays appropriately quickly to be asymptotically (anti)-de Sitter. The remaining requirements of being asymptotically (anti)-de Sitter are easily checked and can be done in a similar way to the examples in~\cite{Ashtekar:1984zz,Ashtekar:2014zfa,Boehmer:2014kxa}. In the case of de Sitter space it is of course important to be careful with coordinates above the cosmological horizon, and this can be dealt with in a rigorous way, again see~\cite{Ashtekar:2014zfa}. It is also possible to observe that the extra contributions to $\omega$ decay as quickly as Kerr-Newman-de Sitter space, which is asymptotically de Sitter.

Thus the general vacuum exterior is asymptotically (anti)-de Sitter, and hence the possibility of matching the slow rotating Garc\'{\i}a metric to this exterior vacuum should be investigated, considering we now have one extra constant available to perform the matching with. We investigate this in the next section.

\section{Matching}
In this section we describe the matching procedure and prove the following result:

{\it The linearised Garc\'{\i}a metric can be matched to an asymptotically (anti)-de Sitter vacuum exterior. }

Now we have already written both the interior and exterior metrics in the form
\begin{align}
ds^2=-\mathcal{A}^2 dt^2+\mathcal{B}^2 dr^2 +r^2[d\theta^2+\sin^2\theta (d\phi-\omega dt)^2].
\end{align}
In what follows we will denote the vacuum region and the interior region by superscripts $(V)$ and $(I)$ respectively. We first apply the following coordinate transformations to the interior fluid region
\begin{align}
t= C_6 t' \quad \varphi=\varphi'+\Omega t
\end{align}
in order to perform the matching. 

The matching conditions up to first order in the angular velocity of the fluid were derived in~\cite{Fodor:2002nt}. Up to this order the matching surface, or the zero pressure surface, remains a sphere,  which we take to be at radius $r=r_1$. Deviations from the spherical symmetry of this surface don't appear until one considers second order corrections to the metric. In order to have a global model in which surface layers of matter are absent, the following Darmois-Israel~\cite{israel} conditions on the metric are required
\begin{align}
g_{ab}^{(V)}&=g_{ab}^{(I)}
\\K_{ab}^{(V)}&=K_{ab}^{(I)}
\end{align}
where $K_{ab}$ is the extrinsic curvature of the metric. These conditions imply the metric functions $\mathcal{A}$ and $\omega$ are $\mathcal{C}^1$ continuous and $\mathcal{B}$ is $\mathcal{C}^0$ continuous at the zero pressure surface $r=r_1$. To ensure absence of surface charges and currents the components $F_{rt}$, $F_{r\varphi}$ and $F_{\theta\varphi}$ of the electromagnetic field tensor are required to be $\mathcal{C}^0$ on this surface.

\subsection{Zero-th order matching}

The zero-th order matching conditions allow us to find the constants $C_6$, $M$, $e$ and the radius $r_1$ of the zero pressure surface in terms of the parameters $\kappa$, $\gamma$ and  $\beta$ of the internal static fluid. Matching of the $g_{\theta\theta}$ component gives
\begin{align}
r_1=\frac{\sin z_1}{\gamma}.
\end{align}
Using continuity of $K_{\theta\theta}$ and $g_{tt}$ we obtain
\begin{align}
C_6=\cos z_1
\end{align}
and continuity of $A_{t,r}$ gives
\begin{align}
e=\frac{\beta}{\gamma}(z_1-\sin z_1 \cos z_1).
\end{align}
The equation of the zero pressure surface, which can equivalently be derived from the junction condition for $g_{tt}$ and $K_{tt}$, lets us solve for the constant $\kappa^2$
\begin{align}
\frac{1}{\kappa^2}=(1-\frac{\Lambda}{\gamma^2})\frac{\tan z_1}{z_1}+2\beta^2(2+z_1 \cot 2z_1).
\end{align}
And finally the mass of the fluid is derived from the continuity of $g_{tt}$
\begin{align}
M&=\frac{r_1}{2}(1-\frac{\cos^2 z_1}{\kappa^2}-\frac{\Lambda}{\gamma^2}-\frac{\Lambda r_1^2}{3})\nonumber\\&+\frac{\beta^2}{2\gamma \sin z_1}(z_1^2+z_1 \sin 2z_1\cos 2z_1+\frac{1}{2}\sin^2 2z_1).
\end{align}

\subsection{First order matching}

Now if we consider the first order matching conditions, we are able to find the values of the first order constants $a$, $\Omega$, $C_1$, $C_2$ and $C_5$ in terms of the parameters of the interior fluid. Continuity of the $g_{t\varphi}$ component of the metric gives the angular velocity of the fluid
\begin{align}
\Omega&= \frac{r_0 \gamma^2}{\sin^2 z_1}(\beta^2 z_1^2(\cot^2 z_1-1)-\beta^2+\kappa^{-2}(1-z_1 \cot z_1))\nonumber\\&+ \frac{1}{\cos z_1}(C_1 \omega_1(r_1)+C_2 \omega_2{r_1})+\frac{2aM \gamma^3}{\sin^3 z_1}-\frac{ae^2 \gamma^4}{\sin^4 z_1},
\end{align}
while continuity of $K_{t\phi}$ allows us to find the Kerr parameter $a$
\begin{align}
a&=\frac{-\sin^5 z_1}{\kappa^2\gamma^4(4e^2\gamma-6M\sin z_1)}(\gamma ^3 r_0 \csc ^2z_1 \sec z_1 (2 \beta ^2 \kappa ^2 z_1^2
   \cot ^3z_1\nonumber\\&+ \cot z_1 (-2 \beta ^2 \kappa ^2 (z_1^2+1)+2 \beta ^2
   \kappa ^2 z_1^2 \csc ^2z_1+3)-2 z_1 (\beta ^2 \kappa ^2+1) \cot
   ^2z_1\nonumber\\&+z_1 (2 \beta ^2 \kappa ^2-\csc ^2z_1))-C_1 \omega_1'(r_1)-C_2 \omega_2'(r_1)). \label{asolution}
\end{align}
Equating magnetic monopole terms gives, as in the $\Lambda=0$ case, we find simply
\begin{align}
C_5=-r_0 e \bar{g}.
\end{align}
Finally the continuity of $F_{\varphi r}$ and $F_{\varphi \theta}$ allow us to solve for both $C_1$ and $C_2$
\begin{align}
C_1&=\frac{4e}{Dr_1^6}(\omega_2'(r_1) \left(5 ae+e  \gamma^2  r_1^2 r_0 \csc^2 z_1 \sec z_1-4 \beta  r_1 r_0(1-z_1 \cot z_1)\right)\nonumber\\&+r_1 \omega_2''(r_1) (ae-\beta  r_1 r_0+\beta  r_1 r_0 z_1 \cot z_1)) \label{c1}
\end{align}
\begin{align}
C_2&=-\frac{4e}{Dr_1^6}
(\omega_1'(r_1)\left(5 ae+e  \gamma^2  r_1^2 r_0 \csc^2 z_1 \sec z_1-4 \beta  r_1 r_0(1-z_1 \cot z_1)\right)\nonumber\\&+r_1 \omega_1''(r_1) (ae-\beta  r_1 r_0+\beta  r_1 r_0 z_1 \cot z_1)) \label{c2}
\end{align}
where the quantity $D$ is defined as
\begin{align}
D&=\omega_1'(r_1) \omega_2''(r_1)-\omega_1''(r_1)
   \omega_2'(r_1).
\end{align}
We can then insert the solution for $a$~(\ref{asolution}) into~(\ref{c1}) and~(\ref{c2}), which will then give two linear simultaneous equations for $C_1$ and $C_2$ which can then easily be solved, although the expressions are lengthy so we omit the details here. This means that all the matching conditions are now satisfied, and therefore our main result is shown to be true. 

\section{Discussion} 

We have shown that the Garc\'{\i}a metric can be matched to an exterior $\Lambda$-vacuum solution up to first order in the fluids angular velocity. This exterior solution is asymptotically (anti)-de Sitter. This contrasts with the $\Lambda=0$ case, where it is the case that the exterior of the slow rotating Garc\'{\i}a metric is not asymptotically flat in general. The addition of a cosmological constant essentially gives rise to an additional degree of freedom by allowing one extra constant in the exterior, meaning there are now enough parameters to match to all of the interior parameters. This allows us to interpret the slow rotating Garc\'{\i}a solution as an isolated body in a cosmological background, an interpretation not possible in the absence of the cosmological term.

This is a very similar result to the one recently obtained in~\cite{Boehmer:2014kxa} for the Wahlquist metric, where to second order in angular velocity the matching to an asymptotically empty domain was found only to be possible once $\Lambda$ was introduced. The addition of charge does not drastically change this result, however the requirement of needing a cosmological constant to perform the matching now appears at first order in the angular velocity rather than second order. In fact with this extra degree of freedom, it can easily be shown, using the same procedure outlined above, that any slow rotating charged perfect fluid interior can be matched to an asymptotically (anti)-de Sitter metric up to first order in angular velocity.

\section*{Acknowledgements}
The author would like to thank Christian G. B\"ohmer for useful comments and discussion.

\end{document}